\documentstyle[12pt,draft]{article}

\begin{document}
\title{Critical and Topological Properties of Cluster Boundaries
in the $3d$ Ising Model}

\author{Vladimir~S.~Dotsenko\cite{landau} and Paul~Windey\\[0.5em]
{\small LPTHE\cite{lpthe}}\\
{\small Universit\'e Pierre et Marie Curie}\\
{\small Bte 126, 4 Place Jussieu}\\
{\small 75252 Paris CEDEX 05, FRANCE}\\[0.5em]
Geoffrey~Harris and Enzo~Marinari\cite{rome}\\[0.5em]
{\small Physics Department and NPAC}\\
{\small Syracuse University}\\
{\small Syracuse, NY 13244, USA}\\ [0.5em]
Emil~Martinec\\[0.5em]
{\small Enrico Fermi Institute and Department of Physics}\\
{\small University of Chicago}\\
{\small Chicago, IL 60637, USA}\\[0.5em]
Marco~Picco\\[0.5em]
{\small Dipartimento di Fisica,}\\
{\small Universit\`a di Roma  {\it Tor Vergata}}\\
{\small Viale della Ricerca Scientifica}\\
{\small 00133 Roma, Italy.}\\[0.5em]}
\date{April 1993}
\maketitle

\vfill
\newpage

\begin{abstract}

We analyze the behavior of the ensemble of surface boundaries of the
critical clusters at $T=T_c$ in the $3d$ Ising model.  We find that
$N_g(A)$, the number of surfaces of given genus $g$ and fixed area
$A$, behaves as $A^{-x(g)}$ $e^{-\mu A}$.  We show that $\mu$ is a
constant independent of $g$ and $x(g)$ is approximately a linear
function of $g$.
The sum of $N_g(A)$ over genus scales as a power of $A$.
We also observe that the volume of the clusters is
proportional to its surface area.  We argue that this behavior is
typical of a branching instability for the surfaces, similar to the
ones found for non-critical string theories with $c > 1$.  We discuss
similar results for the ordinary spin clusters of the $3d$ Ising model
at the minority percolation point and for $3d$ bond percolation.
Finally we check the universality of these critical properties on the
simple cubic lattice and the body centered cubic lattice.

\end{abstract}

\vfill

\begin{flushright}
  {\bf hep-th/9304088}\\
  {\bf EFI 93-24}\\
  {\bf SU-HEP-4240-531}\\
  {\bf LPTHE 93-20}\\
  {\bf ROMA2/93/14}\\
\end{flushright}

\newpage

It is a long-standing hope in theoretical physics that it is possible
to find a formulation of three-dimensional phase transitions dual to
the usual order parameter field theory.  Such a description recasts
the dynamics in terms of fluctuating surfaces.  In the $3d$ Ising
model such a reformulation is indeed possible on the lattice
\cite{string}, and the relevant problem is the existence of a
continuum limit for the surface theory.  Despite much theoretical
effort, real progress in this direction has been slow and difficult.
In order to provide some experimental data to perhaps point in the
right direction, we embarked on a numerical study of surfaces in the
$3d$ Ising model.

The main obstacle to a continuum surface description is the well-known
fingering instability of surfaces in embedding space of dimension
$d>2$\cite{tachyon}.  For $d\le 2$ the theory of random surfaces is
solved\cite{solvers}, providing a striking confirmation of the scaling
predictions of Liouville theory\cite{kpz}.  Another motivation for our
study was to see if the $d>2$ instability plagues Ising surfaces, and
to look for scaling properties in the distribution of surfaces of
fixed genus.  We will present data that shows nontrivial
topology-dependent scaling behavior for self-avoiding surfaces in
$3d$ (for previous work, see \cite{previous}).

Contrary to the widely held belief that a phase transition is
characterized by nested clusters of ordered domains of all possible
sizes, it has been shown \cite{ck} that the distribution of domain
boundaries in the $3d$ Ising model does {\it not} scale at
criticality.  Rather, as the critical point is approached from low
temperatures, islands of flipped spins (which we shall call minority
clusters) merge into a large percolating cluster at a temperature
$T_p$ well below the critical temperature $T_c$.  The dynamics is then
dominated not by the entropy in the cluster distribution but by the
entropy of configurations of the percolating cluster.  Moreover, no
local order parameter of the Ising model reflects this percolation
transition.

However there exists a cluster representation of the Ising model due
to Fortuin and Kasteleyn \cite{fk} which captures the critical
properties of the model, in particular the divergence of the
correlation length is related to the percolation of the FK clusters.
These clusters are formed by adjoining neighboring spins with a
probability $1-e^{-2\beta}$ if they are equal.  The FK representation
led Swendsen and Wang to propose a Monte Carlo that partially defeats
critical slowing down~\cite{sw}. In our simulations we have used the
SW algorithm.  We studied the self-avoiding surfaces bounding both
minority spin clusters and FK clusters.  To explore further the
possible realizations of self-avoidance, we also simulated the
surfaces defined by pure bond percolation \cite{stauffer}.  It should
be emphasized that the `bosonic' surfaces defined by Ising domain
walls are not the same as the `fermionic' surfaces that arise in the
surface reformulation of the lattice Ising model; nevertheless we
expect that they should capture the characteristic features of any
critical surface theory of the $3d$ Ising model.

We ran a medium sized simulation, using roughly four months of time on
RISC workstations. We have analyzed Ising configurations on a $64^3$
body centered cubic (BCC) lattice at a temperature of $\beta = .0857$
using $.3 \times 10^6$ iterations.  We also collected data on simple
cubic lattices of size $32^3$ and $64^3$ at $\beta = 0.221651$
(performing about $6\times 10^6$ and $.25 \times 10^6$ iterations
respectively).  Data were also taken at the minority percolation point
for the Ising model (where we also studied many different lattice
sizes, going from $32^3$ to $100^3$) and for pure bond percolation on
the simple cubic lattice. All of our error analysis has been done by
using jack-knife and binning techniques. For more details see ref.
\cite{uslong}.

On the BCC lattice, we coupled with equal strength both the $6$
nearest and $8$ next-nearest Ising spins so that only three plaquettes
of the dual lattice meet along a dual link.  Since surfaces built this
way on the BCC lattice are naturally self-avoiding, computing the
genus of the dual surface is trivial.  The number of handles $g$ is
obtained through the Euler formula $2-2g=V-L+P$, where $V$, $L$, and
$P$ are the number of vertices, links, and plaquettes, respectively on
the dual surface. On the other hand, the genus definition on the
simple cubic lattice is more problematic, and requires a few choices
(which we discuss in detail in ref. \cite{uslong}) to resolve
ambiguities where surfaces self-touch.

Overall, for the FK clusters we obtained our best results on the BCC
lattice.  We found a scaling law for the number of clusters of volume
$V_{cl}$, $N(V_{cl})\sim V_{cl}^{-\tau}$; the exponent $\tau=2.22$ \cite{wang}
with a large systematic error which could be as high as $3\%$
\cite{uslong}.  We also measured the quantity $A_{cl}$, which counts the
number of cluster sites on the boundary, and found that asymptotically
it was proportional to the cluster volume $V_{cl}$.


Fig.\ \ref{fig1} shows this dependence on the simple cubic lattice.
We see that for very small volumes, the lattice
regularization constrains $V_{cl}$ to equal $A_{cl}$ and for intermediate
volumes, there is a small deviation from linear scaling (as some
interior sites begin to appear).  The plateau that appears around $V_{cl}
= 3000$ indicates the onset of scaling regime where $A_{cl} \propto V_{cl}$.
The growth just at the end of the plot is due to the largest cluster,
which wraps around the lattice and merges with itself to form extra
interior points.  This plateau is the first indication that the
surfaces are not smooth and are unstable towards the formation of
quasi-one-dimensional objects.  The observed proportionality of $V_{cl}$
and $A_{cl}$ is well-known in the context of pure percolation in $2$ and
$3$ dimensions \cite{stauffer}.
We note that in the well understood
case of the $2d$ Ising model (which has a non-pathological continuum
behavior) the cluster perimeter is not proportional to the area
spanned by the cluster.
The area-volume proportionality, along with the relation
\cite{wang} to the magnetization exponent $\delta=1/(\tau-2)$, implies
that the scaling behavior of these `polymers' are related to critical
properties of the Ising model.

Turning to the analysis of the topology of the dual surfaces bounding
the clusters, we consider the distribution $N_g(A)$, where $g$ is the
genus and $A$ is the dual surface area.  In Fig.\ \ref{fig2}, we
present our data for genus $5$ along with a best fit to the functional
form

\begin{equation}
  \protect\label{funform}
  N_g(A) = C_g A^{x(g)} e^{-\mu (g)A}\ .
\end{equation}
The fit is superb.  One of our main results is that this functional
form fits our data very well for $g \geq 2$ up to about $g = 20$ where
our statistics become poor.  The `cosmological constant' $\mu$ is
found to be independent of the genus for $g>2$ (Fig.\ \ref{fig3}).
The value $\mu^{-1}=114\pm 3$ is proportional to the average surface
area (in lattice units) per handle.

In Fig.\ \ref{fig4} we plot the exponent $x(g)$ as a function of the
genus. Once again after a transient region for small genus ($g=0-4$)
we find an almost linear behavior in the region $g=5-15$ with a slope
of $1.25 \pm 0.1$.  The deviations from linearity are small, and in
our observation window they can be fitted with an effective exponent
correction of order $0.1$, or with logarithmic corrections. These make
an estimate of the large genus behavior
of $x(g)$ rather difficult.  The region we are observing
is still transitory, and we cannot exclude that the asymptotic
slope of $x(g)$ could equal one at large $g$.  This value would be
expected if the handles were completely uncorrelated as for a Poisson
distribution $P_g(A)={1\over g!}(\mu A)^g e^{-\mu A}$.

The results presented above have also been analyzed for the simple
cubic lattice. The data agree with those of the BCC and point to a
good universal behavior. The main discrepancies can be traced to the
short distance ambiguities which plague the definition of the dual
surfaces on the simple cubic lattice. The scaling exponents are close
for the two lattices.  The difference among non-universal quantities,
e.g.\ $\mu^{-1}\sim 60$ on the simple cubic lattice, can be understood
from the ratio of the number of plaquettes of the respective
Wigner-Seitz cells.

We will discuss in detail in \cite{uslong} the analogous features for
both the percolation of minority spins of the Ising model and for
non-interacting bond percolation.  The general picture is
interestingly the same; indeed, the slope $d x/d g$ for bond
percolation is compatible with the value for FK clusters.  However,
the slope $dx/dg\sim 0.7\pm 0.1$ in the range $g=3-40$ for Ising
minority spin percolation.  The same caution as before apply to these
slope values.  Another interesting result is the similarity of the
measured area scaling exponents ($\tau=2.18\pm 0.05$ upon
extrapolation to large lattices) as well as the linear relation
between cluster volume and area at the respective critical points (FK
clusters at $T_c$ and minority spin clusters at $T_p$).  The
difference of the $x(g)$ estimated for the two interacting theories we
are studying is interesting. It is possible that in the asymptotic
region one will get $1$ for the slope in both cases, but the large
difference of the measured exponents (which are, on the contrary, very
similar when looking at FK clusters and at bond percolation) says at
least something about finite size corrections, i.e. about the nature
of the interaction. Apart from this effect these results indicate that
the cluster distribution scaling is rather insensitive to the average
cluster density, which is about ten times less for minority spin
clusters. At least partially we may be observing some universality of
different definitions of self-avoiding random surfaces.

We regard the outcome of our topological studies and the behavior
$V_{cl}\sim A_{cl}$ as a strong indication that the cluster boundaries are
in a `branched' phase. The topological evidence suggests that the
surfaces grow fingers which reconnect with a fixed probability per
unit area.  The cosmological constant of the surfaces of fixed genus
is nonzero at the critical point.  Adjusting the temperature away from
criticality will only increase $\mu$, as the large surfaces are
exponentially suppressed (apart from a few surfaces of the size of the
lattice above the percolation threshold).  Therefore there is no
relevant parameter in the theory that could be tuned to allow large
surfaces of low genus.  One can imagine that some additional parameter
(e.g.  one that couples to the Euler density of the lattice surfaces,
which depends on all the spins in a fundamental cell) could be
fine-tuned to multicriticality (for another approach see
\cite{david2}).  Then there would be the possibility to have a scaling
theory at fixed genus.  However, such a tuning will probably not
remove the fingering instability of the surfaces, so we still are
faced with the problem that the continuum theory is not a theory of
surfaces but of quasi-one-dimensional objects.

The approximate linearity of $x$ with $g$ is strongly reminiscent of
the scaling behavior of $d\le2$ random surfaces\footnote{The typical
scaling observed for these $d\le2$ random surfaces is in the free
energy, describing a single fluctuating surface.  Our Ising
configurations, however, consist of a collection of interacting
surfaces.  The scalings we observe are for an ensemble of surfaces and
not the free energy.}.  But note however that similar scaling behavior
can be shown in the `double-scaling' limit of $O(N)$ vector field
theories \cite{vector}, so a similar exponent exists for the scaling
of random graphical networks of fixed topology.

Finally, in order to get a more accurate description of the geometry of
the minority clusters, we analyzed the distribution of
cluster cross sections as a function of their perimeters. Below the {\it
critical} temperature $T_c$
the distribution drops off at a scale of
the order of a few lattice spacings, providing further indication
that the surfaces are composed of small, highly interconnected tubes.
Remarkably, around $T_c$ (recall that the surfaces have
long since percolated at $T_p<T_c$)
we find very good scaling behavior which must be
entirely dominated by the cross sections of the
large percolated cluster. If one were bold
enough to view the percolated cluster as describing the time
evolution of interacting strings, one could speculate that this
behavior is reminiscent of a Hagedorn-type transition.

We would like to thank Stephen Shenker for essential discussions which
led to our investigations.  We are also grateful to NPAC for their
crucial support.  After this work was completed, we received an
interesting paper by M. Caselle, F. Gliozzi and S. Vinti (hep-th
9304001) which puts forward ideas related to those presented here.
This work was supported in part by the Dept. of Energy grants
DEFG02-90ER-40560, DEFG02-85ER-40231, the Mathematical Disciplines
Institute of the Univ. of Chicago, funds from Syracuse Univ., by the
Centre National de la Recherche Scientifique, by INFN and the EC
Science grant SC1*0394.

\vfill
\newpage

\begin{figure}
\caption{\protect\label{fig1}
ln($V_{cl}$/$A_{cl}$) vs. ln($V_{cl}$) for FK clusters on the
$L=64$ SC lattice.}
\end{figure}

\begin{figure}
\caption{\protect\label{fig2}
The number of genus $5$ surfaces as a function of dual
surface area $A$ for FK clusters on the $L=64$ BCC lattice, with a best
fit to the functional form given in equation $1$.}
\end{figure}

\begin{figure}
\caption{\protect\label{fig3}
The dependence of the cosmological constant $\mu(g)$ on genus
for FK clusters on the $L=64$ BCC lattice.}
\end{figure}

\begin{figure}
\caption{\protect\label{fig4}
The dependence of the exponent $x(g)$ (extracted from the
fits to equation \protect\ref{funform}) on genus for FK clusters
on the $L=64$ BCC lattice.}
\end{figure}

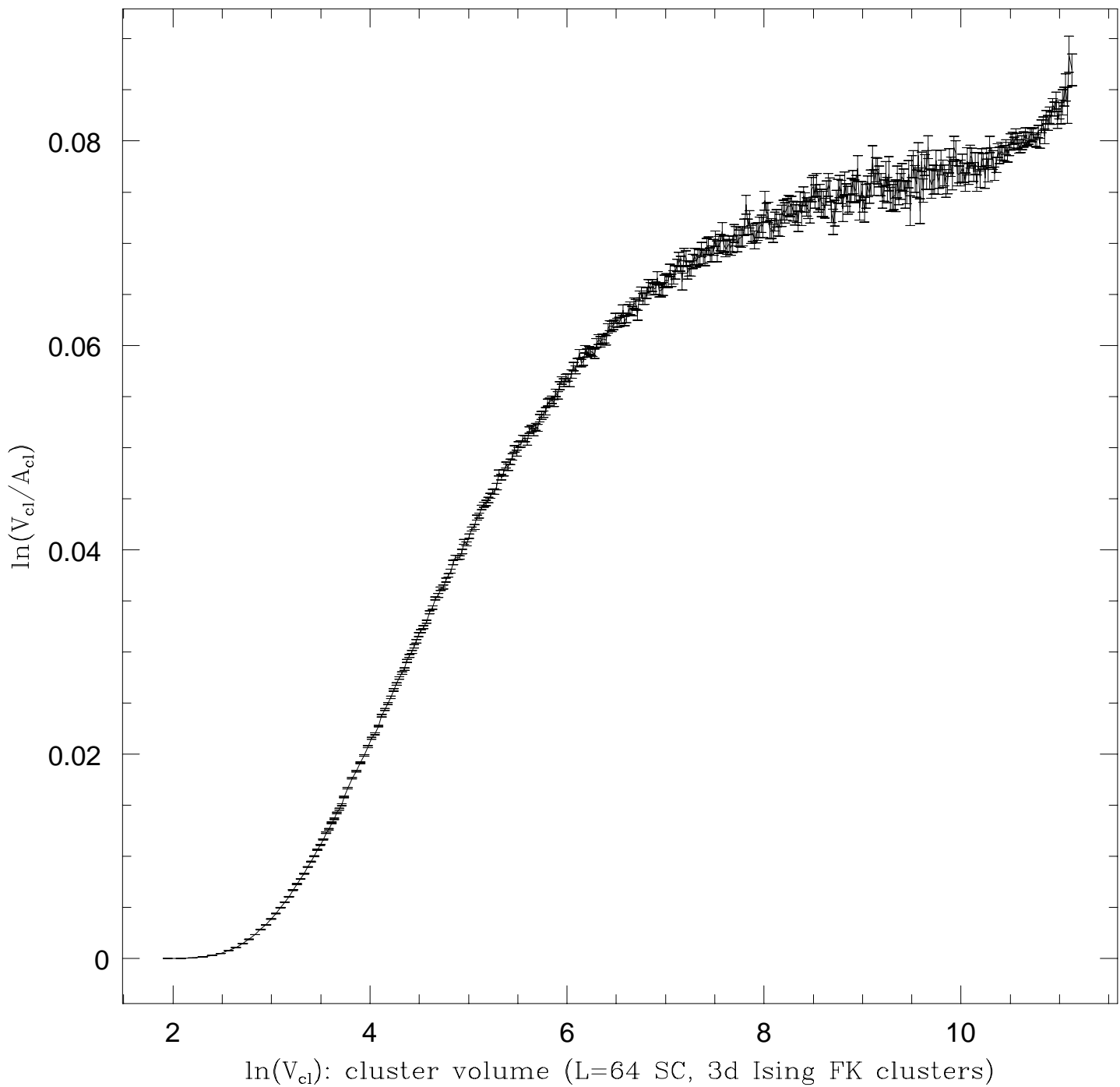
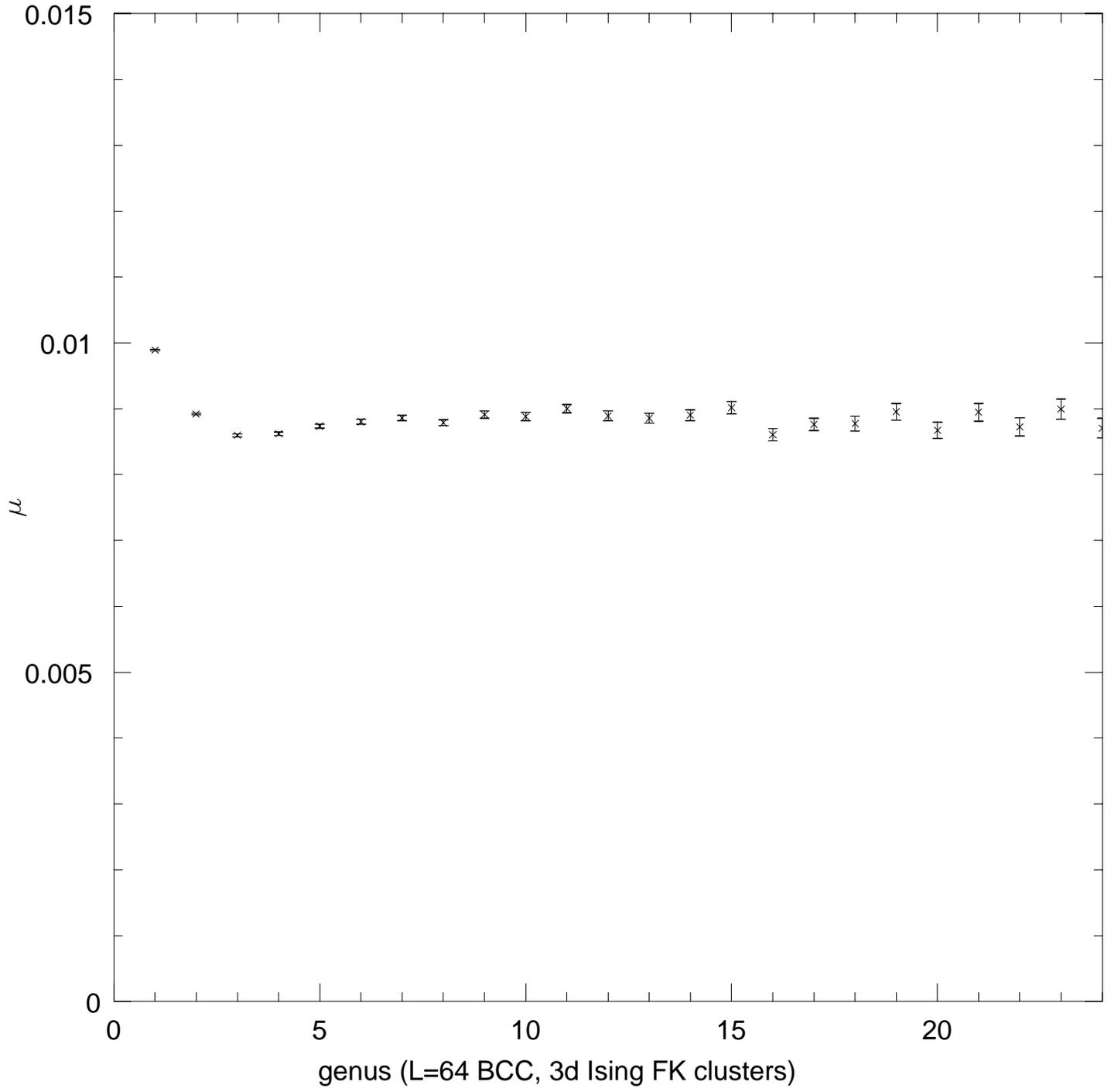
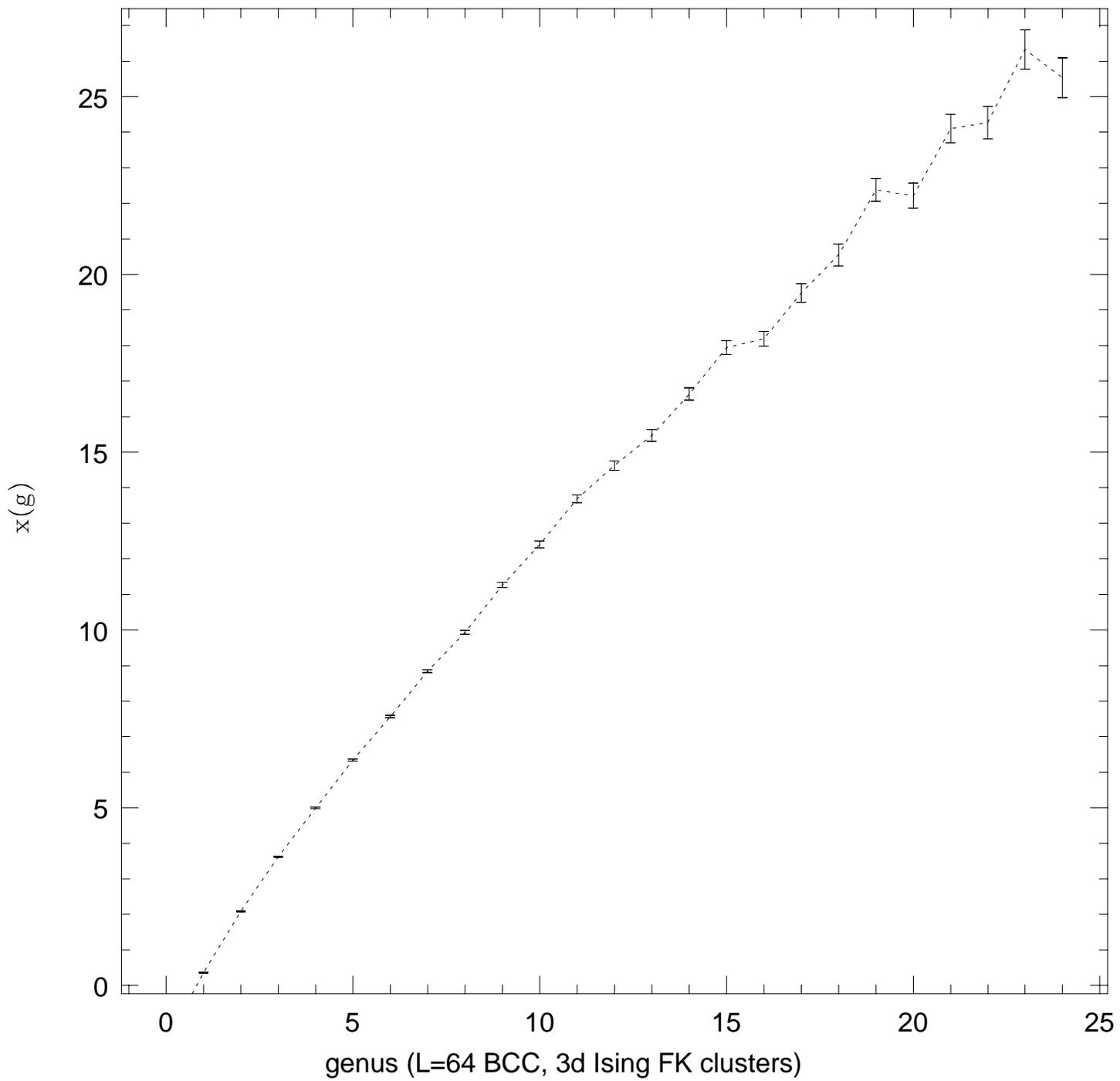
\end{document}